\documentstyle[12pt]{article}
\newcommand{\nc}{\newcommand}

\nc{\dbar}{\bar{\partial}}
\nc{\be}{\begin{equation}}
\nc{\ee}{\end{equation}}

\nc{\beq}{\begin{equation}}
\nc{\eeq}{\end{equation}}
\nc{\bea}{\begin{eqnarray}}
\nc{\eea}{\end{eqnarray}}

\catcode`\@=11

\@addtoreset{equation}{section}
\def\theequation{\thesection\arabic{equation}}

\def\@normalsize{\@setsize\normalsize{15pt}\xiipt\@xiipt
\abovedisplayskip 14pt plus3pt minus3pt%
\belowdisplayskip \abovedisplayskip
\abovedisplayshortskip  \z@ plus3pt%
\belowdisplayshortskip  7pt plus3.5pt minus0pt}
\def\small{\@setsize\small{13.6pt}\xipt\@xipt
\abovedisplayskip 13pt plus3pt minus3pt%
\belowdisplayskip \abovedisplayskip
\abovedisplayshortskip  \z@ plus3pt%
\belowdisplayshortskip  7pt plus3.5pt minus0pt
\def\@listi{\parsep 4.5pt plus 2pt minus 1pt
            \itemsep \parsep
            \topsep 9pt plus 3pt minus 3pt}}

\def\underline#1{\relax\ifmmode\@@underline#1\else
        $\@@underline{\hbox{#1}}$\relax\fi}
\@twosidetrue
\relax

\catcode`@=12

\catcode`\@=11

\def\section{\@startsection{section}{1}{\z@}{3.5ex plus 1ex minus
   .2ex}{2.3ex plus .2ex}{\large\bf}}
\def\thesection{\arabic{section}.}

\def\ps@headings{\def\@oddfoot{}\def\@evenfoot{}
\def\@oddhead{\hbox{}\hfill
        \makebox[.5\textwidth]{\raggedright\ignorespaces --\thepage{}--
        \hfill }}
\def\@evenhead{\@oddhead}
}

\ps@headings

\catcode`\@=12

\relax

\def\figcap{\section*{Figure Captions\markboth
        {FIGURECAPTIONS}{FIGURECAPTIONS}}\list
        {Fig. \arabic{enumi}:\hfill}{\settowidth\labelwidth{Fig. 999:}
        \leftmargin\labelwidth
        \advance\leftmargin\labelsep\usecounter{enumi}}}
 \relax
\def\tablecap{\section*{Table Captions\markboth
        {TABLECAPTIONS}{TABLECAPTIONS}}\list
        {Table \arabic{enumi}:\hfill}{\settowidth\labelwidth{Table 999:}
        \leftmargin\labelwidth
        \advance\leftmargin\labelsep\usecounter{enumi}}}
 \relax
\def\reflist{\section*{References\markboth
        {REFLIST}{REFLIST}}\list
        {[\arabic{enumi}]\hfill}{\settowidth\labelwidth{[999]}
        \leftmargin\labelwidth
        \advance\leftmargin\labelsep\usecounter{enumi}}}
 \relax

\catcode`\@=11

\def\ps@headings{\def\@oddfoot{}\def\@evenfoot{}
\def\@oddhead{\hbox{}\hfill
        \makebox[.5\textwidth]{\raggedright\ignorespaces --\thepage{}--
        \hfill }}
\def\@evenhead{\@oddhead}
}

\ps@headings

\relax

\def\firstpage#1#2#3#4#5#6{
\begin{document}

\begin{titlepage}
\nopagebreak
\title{\begin{flushright}
       \vspace*{-1.8in}
       {\normalsize SISSA-29/97/EP} 
       % -- NUB-#2\\[-9mm]CPTH--RR368.0795}\\[-9mm]
       % {\normalsize hep-th/96}\\[4mm]
\end{flushright}
\vfill
{\large \bf #3}}
\author{\large #4 \\ #5}
\maketitle
\vskip -7mm
\nopagebreak
\begin{abstract}
{\noindent #6}
\end{abstract}
\vfill
\begin{flushleft}
\rule{16.1cm}{0.2mm}\\[-3mm]
\end{flushleft}
\footnotesize{PACS: 11.25Hf, 11.25Mj} \\
\footnotesize{Keywords: BPS, index, type I, $K3$, orientifolds, F-terms}
\thispagestyle{empty}
\end{titlepage}}
\newcommand{\dal}{\raisebox{0.085cm}
{\fbox{\rule{0cm}{0.07cm}\,}}}
\newcommand{\dt}{\partial_{\langle T\rangle}}
\newcommand{\dtbar}{\partial_{\langle\bar{T}\rangle}}
\newcommand{\al}{\alpha^{\prime}}
\newcommand{\mst}{M_{\scriptscriptstyle \!S}}
\newcommand{\mpl}{M_{\scriptscriptstyle \!P}}
\newcommand{\dv}{\int{\rm d}^4x\sqrt{g}}
\newcommand{\lv}{\left\langle}
\newcommand{\rv}{\right\rangle}
\newcommand{\ph}{\varphi}
\newcommand{\sbar}{\,\bar{\! S}}
\newcommand{\xbar}{\,\bar{\! X}}
\newcommand{\fbar}{\,\bar{\! F}}
\newcommand{\zbar}{\,\bar{\! Z}}
\newcommand{\tbar}{\bar{T}}
\newcommand{\ubar}{\bar{U}}
\newcommand{\ybar}{\bar{Y}}
\newcommand{\phb}{\bar{\varphi}}
\newcommand{\cm}{Commun.\ Math.\ Phys.~}
\newcommand{\pr}{Phys.\ Rev.\ D~}
\newcommand{\prl}{Phys.\ Rev.\ Lett.~}
\newcommand{\pl}{Phys.\ Lett.\ B~}
\newcommand{\ibar}{\bar{\imath}}
\newcommand{\jbar}{\bar{\jmath}}
\newcommand{\np}{Nucl.\ Phys.\ B~}
\newcommand{\e}{{\rm e}}
\newcommand{\gsi}{\,\raisebox{-0.13cm}{$\stackrel{\textstyle
>}{\textstyle\sim}$}\,}
\newcommand{\lsi}{\,\raisebox{-0.13cm}{$\stackrel{\textstyle
<}{\textstyle\sim}$}\,}
\date{}
\firstpage{95/XX}{3122}
{\large\sc\bf BPS states and supersymmetric index\\ in $N=2$ type 
I string vacua} {Jose F. Morales$^{a}$ and Marco 
Serone$^{a,b}$}%\\[-3mm]
{\normalsize\sl $^{a}$International School for Advanced Studies, ISAS-SISSA
\\[-3mm]
\normalsize\sl Via Beirut n. 2-4, 34013 Trieste, Italy\\[-3mm]
\normalsize\sl \\[-3mm]
\normalsize\sl 
$^{b}$Istituto Nazionale di Fisica Nucleare, sez.\ di Trieste, Italy\\[-3mm]
\normalsize\sl e-mail:morales,serone@sissa.it \\[-3mm]}
%\normalsize\sl $^c$SISSA, I-34100 Trieste, Italy\\[-3mm]}
{We study the moduli dependence of a class 
of couplings in $K3\times T^2$ compactifications 
of type I string theory, for which one-loop
amplitudes can be written in terms of an $N=2$
supersymmetric index. This index is determined for
generic models as a function of the BPS spectrum. 
As an application we compute the one-loop moduli dependence of the 
$F_gW^{2g}$ couplings, where $W$ is the $N=2$ gravitational 
superfield, for type I compactifications based on the 
Gimon-Johnson $K3$ orientifolds, showing 
explicitly their dependence on the aforementioned index.} 

\section{Introduction}

During the recent developments achieved in string and 
quantum field theories with extended
supersymmetry, it has become evident that a prominent
role in governing the dynamics is played by BPS states.
These special states provide important clues in the exploration 
of the strong coupling regimes of the corresponding theories.
The appeareance of solitonic massless BPS states in $N=2$ 
supersymmetric Yang-Mills theories \cite{sw},
the prediction of U-duality in type II string theories of the
existence of BPS states carrying Ramond-Ramond charges \cite{ht},
subsequently identified with D-branes \cite{pol}, and the resolution
of the conifold singularity via a solitonic charged BPS state \cite{st}
are only few examples of the vast number of results supporting 
their importance.\\ 
An interesting connection between BPS states and generalized Kac-Moody 
currents has been found in \cite{hm}. In this reference the authors
study threshold and gravitational corrections in $N=2$ four-dimensional
compactifications of the heterotic string. These corrections can 
be written in terms of the 
$N=2$ supersymmetric index defined in \cite{vc} and are determined
purely in terms of BPS states. In particular the index is shown to count 
the difference between the number of hyper and vectormultiplets in the
effective four-dimensional theory.   
The scope of this paper is to show
that an analogous result holds in $N=2$ type I string compactifications.
The $N=2$ supersymmetric index is realized in type I compactifications 
and then related to the corresponding threshold and gravitational 
corrections. The connection between type I amplitudes and this index
was first pointed out by Antoniadis et al. in ref.\cite{abf}.
For a generic compactification we compute this index as a function of
the BPS spectrum, finding again that BPS contributions enter only through
the difference between hyper and vectormultiplets \footnote{Strictly
speaking this is not the case for the $F_1$-coupling, see below.}.  
As an application of these results we will 
explicitly compute one-loop amplitudes of higher derivative F-terms 
$F_g(X)W^{2g}$, with $W$ the $N=2$ gravitational superfield and $X$ a 
generic chiral vector superfield, for a class of type I string 
compactifications obtained as an orientifold of type IIB theory 
\cite{bia,gj,aj,ang},
generalizing the computation performed in \cite{ms} for the 
${\bf Z}_2$ model \cite{gp,pra} and pointing out
explicitly the BPS dependence of these amplitudes. 
This tower of gravitational couplings has been extensively studied in 
the context of string dualities. In particular for type II strings
they are given by topological partition functions in twisted Calabi-Yau
sigma models \cite{bcv}, while in heterotic compactifications they can
be extracted from a simple one-loop amplitude \cite{agnt,mos}. The type I
computation of these terms, like in the heterotic case, is given by a
one-loop amplitude which can be written in terms of the mentioned 
supersymmmetric index. This index encodes the compactification
model-dependence of these couplings, while an universal part coming 
from the spacetime correlations completes its structure.   

The paper is organized as follows.
In the next section we realize the
supersymmetric index in $N=2$ type I compactifications and compute it for
a generic model as a function of its four dimensional BPS spectrum. 
In section three we briefly review the construction of $K3$ orientifolds
given by \cite{gj} and then we compute the $F_g$ couplings in these vacua.
In the two appendices we report some technical details needed to compute 
the correlation function and the values of the indices for the various 
models. Some conclusions are given in the last section. 

\section{BPS states and the $N=2$ supersymmetric index} 

In this paper we study the moduli dependence of one-loop amplitudes of
couplings in $N=2$ supersymmetric four-dimensional lagrangians arising as
$K3\times T^2$ compactifications of type I string theory. We will
consider in detail the tower of $F_g$ gravitational couplings, although in 
this section we will restrict our analysis to the main feautures of the
involved amplitudes in order to allow a straightforward
generalization to a wider 
class of terms. In particular we discuss some examples of couplings in
which one-loop corrections can be written in terms of the ``new supersymmetric
index'' \cite{vc} realized in $N=2$ type I compactifications.
The important point in the considered amplitudes is that the internal
theory, entering simply through an index, receive contributions 
only from its ground states. As it will be explicitly shown in our 
computations, the space-time part of the correlation function can be 
reduced to a supersymmetric partition function in the odd-spin structure, where
a cancellation between fermion and boson determinants holds and again
only ground states give a non-vanishing result. 
In this way we reduce the amplitude to a sum of BPS states contributions.\\ 
One loop amplitudes in string computations involve in general a sum over 
the spin structures carried by the involved surfaces.
Using the Riemann identity we can relate the sum over spin structures
to an odd spin structure correlation of new operators obtained
by a triality rotation of the original ones 
\cite{gjn}. In the last section we use this procedure to reduce the 
relevant amplitude for the $F_g$ couplings (a bunch of gravitons and 
graviphotons) to
an odd spin structure correlation. Other physical one-loop
corrections as thresholds to gauge
couplings are also determined from odd spin structure computations \cite{agn}.
In this section we will see how these odd correlations are related 
to a realization of the
mentioned supersymmetric index. 

\subsection{One-loop corrections in type I string compactifications}

In order to understand better the general structure of the kind of 
amplitudes involved in our discussion, let us briefly work it out two simple
examples of relevant odd spin structure correlations in $K3\times T^2$
type I compactifications.
\vspace{0.3 cm}

a) {\em $F_1$ gravitational coupling}
\vspace{0.3 cm}

In this first example we discuss the moduli dependence of the $R^2$
coupling in the effective four-dimensional action, with
$R_{\mu\nu\rho\lambda}$ the Riemann tensor. This can be extracted 
\cite{agn} from the on-shell three-point function of two gravitons 
and a T-modulus computed in the odd spin structure.
Type I computations involve also a sum over the 
surfaces that we will denote respectively by T, K, A, M for the torus, 
Klein bottle, annulus and M\"obius
strip. The torus contribution can be written as
\bea
\epsilon^{\mu\nu\lambda\rho}p_{1\lambda}p_{2\rho}
p_2^{\alpha}p_1^{\beta}\:\Theta^{(T)}_T~&=&~
\int_{\Gamma}\frac{d^2\tau}{\tau_2}
\int \prod_{i=1}^{3} d^2z_i 
\langle V_h^{\mu\alpha}(p_1,z_1)\, V_h^{\nu\beta}(p_2,z_2)\,\nonumber\\
&\times&
V^{(-1,-1)}_T(p_3,z_3)\,
T_F(z)\,\tilde{T}_F(\bar{z})\rangle_{\rm odd}
\label{I}
\nonumber
\eea
where $\tau = \tau_1 + i\tau_2$ is the modular parameter of the
world-sheet torus, ${\Gamma}$ its fundamental domain and 
$V^{(-1,-1)}_T$ is the $T$-vertex
operator in the $(-1,-1)$ left-right ghost picture
\beq
V^{(-1,-1)}_T(p) = \!e^{-(\phi+\tilde{\phi})}\Psi_+(z,\bar{z})
\,e^{ip\cdot \!X}\!\nonumber\label{VT}
\eeq
$\Psi_+(z,\bar{z})$ being a primary field of dimension $(1/2,1/2)$.
The $(-1,-1)$ ghost picture takes into account 
the left-right Killing spinors on the world-sheet torus
while the supercurrent insertions ($T_F=T_F^+ + T_F^- +$space-time
part, and the same for the left one $\tilde{T}_F$) 
soak up the gravitino zero modes. The $V_h$'s represent the gravitons, 
whose vertices are given by:
\beq
V_h^{\mu\nu}(p) = (\partial X^{\mu}+ip\!\cdot\!\psi\,\psi^{\mu}) 
(\bar{\partial} X^{\nu}+ip\!\cdot\!\tilde{\psi}\,\tilde{\psi}^{\nu})\, 
e^{ip\cdot \!X} \label{Vh} \eeq
The analysis of this amplitude is just a left-right symmetric version
of the one performed in \cite{agn}.
The OPE of the $N=2$ internal superconformal algebras are given by
\beq
\tilde{T}_F^{\mp}(\bar{w})T_F^{\mp}(w)\Psi_{\pm}(z,\bar{z}) = \mp \tilde{J}(\bar{w})J(w)
\Phi_{\pm}(z,\bar{z}) + ...
\label{tps}
\eeq
where $J,\tilde{J}$ are the U(1) currents associated to the $N=2$ 
superconformal 
algebras, $\Phi_{\pm}$ is the dimension (1,1) upper component of $\Psi_{\pm}$
and ... are contour integrals that give vanishing contribution to 
this amplitude.
We can then relate 
$\Theta_T$ to a $T$-derivative of a quantity $\Delta$ written as
$$
\Theta^{(T)}_{T}~=~i\partial_T\Delta^{(T)}
$$ with
\beq
\Delta^{(T)}~=~-i
\int_{\Gamma}\frac{d^2\tau}{\tau_2}\,C_{T}\eeq
and 
\beq
C_{T}\equiv {\rm Tr}_{RR}(-1)^{F_L+F_R}\,F_L F_R 
\,q^{\Delta}\bar{q}^{\bar{\Delta}}
\label{ind1} \eeq
where $q=e^{2\pi i \tau}$, the trace is restricted to the 
Ramond-Ramond sector and contains the momenta
lattice sum in the torus direction, and 
$\Delta$ is the conformal dimension of the
state propagating around the loop. $F_L,F_R$ are the fermionic numbers,
i.e. the zero modes of the $U(1)$ currents $J_{L},J_{R}$,
which soak up the four zero modes of the free fermions associated
to the torus direction and then are necessary to get a non-vanishing result.  
The space-time part contributes only through the eight zero modes
needed in the torus for the odd-spin structure. 
In the last section we will study the whole tower of $F_g$ couplings,
which includes, besides the two gravitons,   
an additional bunch of $(2g-2)$ spacetime operators, obtained
from the graviphotons through a triality rotation induced by the
spin-structure sum. The internal structure is therefore untouched
and only the space-time part of the amplitude will be modified.
In this section we will concentrate in this
internal part $C_T$ (and similar quantities for the
rest of the surfaces) which realizes the $N=2$ supersymmetric 
index. The contributions given by the other surfaces can be analyzed in an 
analogous way. Proceeding along the same lines followed for the torus, we
relate the amplitudes to an integration
in the corresponding worldsheet moduli of a spacetime 
correlation and an internal contribution through an index written as:
\bea
C_K &\equiv& {\rm Tr}_{RR}\frac{(F_L+F_R)}{2}\,(-)^{F_L+F_R}\, \Omega  
\,q^{\Delta}\bar{q}^{\bar{\Delta}}
\nonumber\\
C_A&\equiv& {\rm Tr}_{R}(-1)^F \,F\, 
q^{\Delta}\bar{q}^{\bar{\Delta}} \label{ind2} \\
C_M&\equiv&{\rm Tr}_{R}(-1)^F\, F\, \Omega\, 
q^{\Delta}\bar{q}^{\bar{\Delta}} \nonumber
\eea
Again the $F$ insertions provide the correct number of zero modes
(two in this case) that we need to soak up in order to get a non-vanishing
result. 

\vspace{0.3 cm}
b) {\em Threshold corrections to gauge couplings}
\vspace{0.3 cm}

The second example refers to moduli dependence of one-loop threshold
corrections to gauge couplings. More general corrections of this kind
have been studied in \cite{bach}. We restrict ourselves to show the
connection of these amplitude with the considered index.
As before, we can extract the moduli dependence from 
a three-point function where now, instead of gravitons, we insert the gauge
field vertex operators:
\beq
V_A^{\mu,a}(p,z) = 
(\partial_{\tau} X^{\mu}+ip_{\tau}\!\cdot\!\psi\,\psi^{\mu}) 
e^{ip\cdot \!X}\!\lambda^a
\label{Va}
\eeq
with ``a'' an index in the adjoint of the gauge 
group and $\lambda^a$ the
corresponding Chan-Paton matrix. 
The relevant amplitude is then given by
\bea\hspace{-1cm}\epsilon^{\mu\nu\lambda\rho}p_{1\lambda}p_{2\rho}
\delta^{ab}\:\Theta^{A,M}_{T}\! ~=~\!
\int \frac{dt}{t}\!
\int\prod_{i=1}^{2} dt_i\,d^2z
\langle V_A^{\mu,a}(p_1,t_1)\, V_A^{\nu,b}(p_2,t_2)\,
V^{(-1)}_T(p_3,z)\, T_F(z_0)\rangle^{A,M}_{\rm odd} \label{I2} \eea
where $V_T^{-1}$ is the closed modulus
\beq
V_T^{(-1)}(p)= (e^{-\phi}\Psi(z,\bar{z})+e^{-\tilde{\phi}}
\tilde{\Psi}(\bar{z},z))e^{ip\cdot X} \eeq
where $\Psi$ and $\tilde{\Psi}$ are respectively the components 
of dimensions (1/2,1) and (1,1/2) of an $N=2$ superfield 
and $\hat{T_F}\equiv T_F+\tilde{T}_F$ is the 
left-right symmetric picture changing operator.

The four spacetime zero modes required in the odd spin structure come
from the fermion part of the gauge field vertices reproducing the correct 
kinematic factor in (\ref{I2}). As before the $N=2$ OPE allows us to write
the internal contribution as a $T$-derivative of a trace in the Ramond sector
$$ \Theta^A_{T}~=~i\partial_T\Delta^A_a $$
with \beq \Delta^A_a~\sim~\int\frac{dt}{t} C_A^a \eeq
and 
\beq C_A^a\equiv {\rm Tr}_{R}(-1)^{F}F\,Q_a^2q^{\Delta}
\bar{q}^{\bar{\Delta}}\label{ind3} \eeq
where $Q^2_a$ is the charge of the state propagating around the loop.
The M\"obius strip contribution on the other hand is just given by an 
$\Omega$ insertion in this trace. We recognize again the indices found
in the previous example.
In this way we have expressed some one-loop corrections to 
four-dimensional effective actions arising from compactifications
of type I strings, in terms of a realization of
the $N=2$ supersymmetric index in these theories. 

\subsection{ BPS states and $N=2$ Supersymmetric indices }
In the beginning of this section we argued that the quantities
(\ref{ind1},\ref{ind2}) can be written as a sum over BPS 
contributions. This subsection is devoted to determine them in terms of 
this spectrum for a generic $K3\times T^2$ type I 
compactifications. 
Threshold and gravitational one-loop corrections for analogous
compactifications of the heterotic string are also written in terms of
this $N=2$ supersymmetric index. Exploiting the representation
properties of the internal superconformal algebra for these compactifications,
Harvey and Moore \cite{hm} found that the index counts
the difference between the number of 
BPS hyper and vectormultiplets at each level of mass. 
We will follow the
lines of this reference to find similar results for the relevant indices
involved in type I compactifications. The analysis is performed for the
case in which no Wilson lines are turned on. Our considerations are however
general, and the modifications brought by their inclusions will be 
pointed out.
 
The internal superconformal theory associated to $K3\times T^2$
compactifications 
of type I theory is a sum of two pieces, corresponding to the open and closed string 
sectors. The open sector is realized with a $(c=3,N=2)\oplus (c=6,N=4)$ SCFT while the 
closed sector is associated to the conformal theory that arises after an $\Omega$-
projection of the $[(c=3,N=2)\oplus (c=6,N=4)]_L\otimes [(\tilde{c}=3,N=2)
\oplus
 (\tilde{c}=6,N=4)]_R$ SCFT, where $\Omega$ is the world-sheet parity operator.
In order to relate (\ref{ind1},\ref{ind2}) 
to a counting of four-dimensional BPS states
let us review the structure and superconformal content of these
states in type I compactifications.

String states in the open sector satisfy the mass condition (in the
Neveu-Schwarz sector): 
\be \frac{1}{2}M^2=\frac{1}{2}p^2+(N-\frac{1}{2})+h_{int} \ee
with $N$ the oscillator number associated to the space-time and 
torus directions,
$h_{int}$ the conformal weight in the $N=4$ SCFT and $p$ the Kaluza-Klein 
momentum
coming from the torus. The BPS bound $M^2=p^2$ is then satisfied only
by the six-dimensional massless Neveu-Schwarz states $(N=1/2,h_{int}=0)$
and $(N=0,h_{int}=1/2)$, which after a further torus compactification
generate all the four dimensional vector and
hyper BPS multiplets. Note that this is not what happens in $N=2$ heterotic models
where the BPS condition (in the 
Neveu-Schwarz sector) reads
\bea
\frac{1}{8}M^2=\frac{1}{2}p_R^2=\frac{1}{2}p_L^2+(h-1)
\eea
with $p_L,p_R$, the torus momenta including winding and Kaluza-Klein modes,
and $h$ the conformal weight of the states in the $c=26$ CFT.   
In this case, the number of BPS states depends on the level of mass, since
for a fixed $p_R^2$ each point in the lattice $p^2_R-p^2_L\in 2{\bf Z}>0$ 
defines additional BPS states with $h>1$ besides the
six-dimensional massless one $h=1$. 

The structure of the type I open Ramond sector is simply
obtained by spectral flow of the two massless Neveu-Schwarz
representations seen before. 
The conformal content is displayed in the following table:
 
{\bf Sector}\hfill{\bf Vectormultiplets}\hfill{\bf Hypermultiplets}
\hspace*{1.0 cm}\\
{\rm Ramond}\hfill $(1/8,\pm 1/2)\otimes(1/4,1/2)$ \hfill 
$2\times(1/8,\pm 1/2)\otimes (1/4,0)$ 

where, following the notation of \cite{hm}, we denote with $(h,q)\otimes (h^{\prime},I)$
a state with conformal weight $h$ and U(1) charge $q$ of the $c=3,N=2$ theory and
weight $h^{\prime}$ and representation $I$ of the SU(2) current of the $c=6,N=4$
theory.\\ 
We are now ready to compute the indices (\ref{ind2}) associated 
to the open string sector.
The fermionic numbers decompose
as $F=F^{(1)}+F^{(2)}$, $F^{(1)}$ and $F^{(2)}$ being
the zero modes of the U(1) currents 
$J^{(1)}$ and $J^{(2)}$ associated to the $N=2$ and $N=4$
superconformal algebras respectively ($J^{(2)}= 2 J^3$, where $J^3$ is
the Cartan 
element of the $N=4$ SU(2) current).   
Since in $SU(2)$ representations the eigenvalues of $J^3$ come always 
in pairs, the only non-vanishing contribution to the indices
come from the $F^{(1)}$ insertion. Equivalently, only the $F^{(1)}$
insertion, which soaks up the zero modes of the free fermions associated to 
the $T^2$ torus, give a non-vanishing result.  
We are then left with the trace
\be 
C_{\rm A}+C_{\rm M}=2\,{\rm Tr}_{N=2}F^{(1)}(-)^{F^{(1)}}
{\rm Tr}_{N=4}(-)^{F^{(2)}}q^{\Delta}, 
\label{ind4}\\ 
\ee
where the trace now runs over the $\Omega$-invariant states
\footnote{Note that the $\Omega$-projection in the open sector gives
constraints only on the Chan-Paton degrees of freedom.}, including
the Chan-Paton degrees of freedom.  
The $N=2$ part is common to both multiplets, while the $N=4$ SCA 
enters only through the Witten indices \cite{te}: 
\bea
{\rm Tr}_{(1/4,0)}(-)^{F^{(2)}}&=&1\nonumber\\
{\rm Tr}_{(1/4,1/2)}(-)^{F^{(2)}}&=&-2
\eea 
Finally we find
\begin{eqnarray} C_{\rm A}+C_{\rm M}&=&2\sum_{p\in\Gamma}
(\frac{1}{2}e^{i\frac{\pi}{2}}-\frac{1}{2}e^{-i\frac{\pi}{2}})({\rm Tr}_{I=0}
(-)^{F^{(2)}}+{\rm Tr}_{I=1/2}(-)^{F^{(2)}})e^{-\pi t|p|^2}= \nonumber \\
&=&4i(n_H^{open}-n_V^{open})\sum_{p\in\Gamma}
e^{-\pi t|p|^2} \end{eqnarray}
where $\Gamma$ represents the lattice momenta sum and $n_V^{open}$, $n_H^{open}$ 
are respectively the
number of massless four dimensional vector and hypermultiplets in the open
string sector. 
We have extracted the overall factor
$(n^{open}_H-n^{open}_V)$, corresponding to the common degeneration 
to all levels of mass of the BPS number, as was already discussed 
before. More general backgrounds including Wilson lines on the torus
can be analyzed. In this case, the $T^2$ momenta lattice will be shifted
by the included gauge field expectation values, and a given number of 
massless vector and hypermultiplets will get masses. 
Notice, however, that their difference will be left invariant, since
this Higgs mechanism will always give mass to an hyper-vector pair.    

Let us now turn to the closed string spectrum. The bosonic BPS content
in this sector arises from a tensor product of the aforementioned
NS-NS (R-R) massless
representations symmetrized (antisymmetrized) under $\Omega$.
Let us first notice that for each R-R ground state in the $N=4$
SCFT we can construct four states, taking into account the multiplicities
coming from the $N=2$ space-time and torus algebra representations. 
Altough the trace runs only in the R-R sector, using spectral flow, we can see that
all the bosonic content of BPS multiplets is taken into account, since $\Omega$-even contributions
to this trace count the NS-NS states kept by the $\Omega$-projection.
The counting of R-R ground states in the $N=4$ theory is simply 
given by the geometrical structure of $K3$: 20 states 
$(1/4,0)\otimes (1/4,0)$,
corresponding to the $h^{1,1}=20$
cohomologically distinct $(1,1)$ differential forms on $K3$ 
and one multiplet $(1/4,1/2)\otimes (1/4,1/2)$ counting the four forms
given by $h^{0.0},h^{2,0},
h^{0,2},h^{2,2}$. As is clear, there are no states 
$(1/4,1/2)\otimes (1/4,0)$ or 
viceversa because $h^{1,0}=h^{0,1}=h^{2,1}=h^{1,2}=0$.
We can finally write the torus index as
\be
C_{\rm T}={\rm Tr}_{RR}^{N=2\oplus
N=4}F_LF_R(-)^{F_L+F_R}
q^{\Delta}\bar{q}^{\bar{\Delta}}=-
(n_H^{closed}+n_V^{closed})
\sum_{p\in\Gamma}
e^{-\pi\tau_2|p|^2} 
\ee
where $n_H^{closed}+n_V^{closed}$ are the 24 massless four
dimensional hyper and vectormultiplets, corresponding to the
Witten index 
\cite{witt} ${\rm Tr}_{RR}(-)^{F_L^{(2)}+F_R^{(2)}}=\chi(K3)$.
The $N=2$ part 
enters only through the lattice sum and the overall $(-1)$ factor.

The last involved quantity is the index related to the Klein Bottle.
If we call $\Omega_{int}$ 
the worldsheet parity operator restricted to the $K3$ part,
we can observe that vectors, being constructed
from left-right symmmetric spacetime+torus combinations of states,
are counted by $\Omega_{int}$ with a minus sign in the $RR$ sector. 
This observation allows us to write finally: 
\be
C_{\rm K}={\rm Tr}_{RR}^{N=2\oplus
N=4}\frac{F_L+F_R}{2}(-)^{F_L+F_R}\Omega
q^{\Delta+\bar{\Delta}}=i(n_H^{closed}-n_V^{closed})
\sum_{p\in\Gamma}e^{-\pi t|p|^2} \ee
where again the overall factor (i) comes from the $N=2$ part.

We achieved the final goal of this section. The contributions associated to the
Klein bottle, annulus and M\"obius strip depend only on the difference between
the number of hyper and vector BPS states of a given compactification. As 
noted in \cite{hm}, this dependence ensures the smoothness of amplitudes 
in the moduli space,
since BPS states in $N=2$ theories always appear and disappear in 
hypermultiplet-vectormultiplet pairs. Notice, however, that 
$C_{\rm T}$ enters in a different way, but being 
associated to corrections in 
$N=4$ supersymmetric theories, it does not contribute in general. 
Among the couplings we analyzed, it gives a 
non-trivial contribution
only to the gravitational coupling $F_1$, as we have seen before and we will see
in greater detail in the next section.

\section{$F_{g}$ terms on $K3$ orientifolds}
As an application of the results found in the last section, we want
explicitly compute in the following the gravitational couplings $F_g$
at 1-loop for $K3$ orientifold models, showing their BPS
dependence.

\subsection{Review of $K3$ orientifolds}

In this subsection we briefly review the construction of orientifolds of
Type IIB string theory\cite{bia,gj,aj,ang}, following in particular the
construction given by  \cite{gj}. An orientifold \cite{shd}
is a generalization of an orbifold where the discrete group by which we 
mod out contains in general also the world-sheet parity operator $\Omega$.
We will consider in particular orientifolds of $K3$, in its orbifold limits
$T^4/{\bf Z}_N$ with $N=2,3,4,6$. As discussed in \cite{gj}, there is a 
freedom in choosing the orientifold group; closure under group 
multiplication allows two choices:
\be {\bf Z}_N^A=\{1,\Omega,\alpha_N^k,\Omega_j\} \ \ \ k,j=1,2,...,N-1; \ \ 
(N=2,3,4,6) \ee and \be {\bf Z}_N^B=\{1,\alpha_N^{2k-2},\Omega_{2j-1}\} 
\ \ \ k,j=1,2,...,N/2; \ \ (N=4,6) \ee
where, following the same notation of \cite{gj}, $\alpha_N$ is the 
generator of the discrete group ${\bf Z}_N$ and 
$\Omega_j\equiv\Omega\cdot\alpha_N^j$. The cancellation of tadpole 
divergencies will require in general the addition of sources of R-R charges,
i.e. of D-branes, with the corresponding open string sectors. It has been 
found in \cite{gj} that for the A-models we always need 32 D-9 branes and 
32 D-5 branes, excluding the case of ${\bf Z}_3^A$ where we do not have 
D-5 branes at all. On the other hand the ${\bf Z}_4^B$ model does not have 
open sectors at all, while the ${\bf Z}_6^B$ has only 32 D-5 
branes\footnote{Strictly speaking we are not considering here the number of 
dynamical D-branes in the model, but the number of values the Chan-Paton
factors take.}. Since, as we will show in the next subsection, the amplitudes
we want to compute are invariant for small perturbations of the models, like
moving the D-5 branes or varying vacuum expectation values continuosly,
we only consider here the massless spectrum of these models, in the 
case of maximum gauge group, when all the D-5 branes
are overlapped on a fixed point of the $K3$ orbifold.

As was argued in section two only the six-dimensional massless
states are relevant to the indices computation. This spectrum is reported in
\cite{gj} for the different orientifold models which we are considering.
It always contains a universal part from the closed string sector formed 
by the gravitational multiplet and one tensormultiplet\footnote{Note that in 
\cite{ang} it has been found, among others, a model without tensormultiplets.}. 
On the other hand the open sector provides the gauge group and 
charged matter in a given representation. 
It has been shown in detail in \cite{blp} for the ${\bf Z}_2^A$ model
that potentially dangerous U(1) anomalies, in general presented in these 
models, are simply removed by a Higgs mechanism that gives mass to the  
corresponding U(1) gauge field. Our amplitudes, however, being proportional
to the supersymmetric index, do not depend on this Higgs phenomenon
that give mass to an hyper-vector multiplet pair. In the 
following we will then simply ignore it.
The four dimensional models we want to consider are obtained by a further
compactification on a $T^2$ torus, giving the same hypermultiplet 
massless spectrum and $n_T+3$ further vectormultiplets coming from the 
closed string sector, where $n_T$ is the number of tensormultiplets in 
six dimensions.

\subsection{Computation of the $F_g$ couplings}

The relevant amplitude we consider here to compute the $F_g$ 
couplings involves two gravitons and $2g-2$ graviphotons whose vertex 
operators are:
\begin{eqnarray} V^{\mu\nu}_g(p)&=&(\partial X^{\mu}+ip\cdot\psi\psi^{\mu})
(\bar{\partial}X^{\nu}+ip\cdot\tilde{\psi}\tilde{\psi}^{\nu})e^{ip\cdot X} \nonumber \\
V_{\gamma}(p)&=&(Q_1^{(L)}+Q_1^{(R)})(Q_2^{(L)}+Q_2^{(R)})V_g(p)\end{eqnarray}
where
\begin{eqnarray}
Q_{\alpha,1}^{(L)}+Q_{\alpha,1}^{(R)}&=&\oint dz
e^{-\frac{\phi}{2}}
S_{\alpha}\Sigma e^{i\frac{H_5}{2}}(z)+\oint d\bar{z}
e^{-\frac{\tilde{\phi}}{2}}\tilde{S}_{\alpha}\tilde{\Sigma}
e^{i\frac{\tilde{H}_5}{2}}(\bar{z}) \nonumber \\
Q_{\alpha,2}^{(L)}+Q_{\alpha,2}^{(R)}&=&\oint dz
e^{-\frac{\phi}{2}}
S_{\alpha}\bar{\Sigma}e^{i\frac{H_5}{2}}(z)+\oint d\bar{z}
e^{-\frac{\tilde{\phi}}{2}}\tilde{S}_{\alpha}\tilde{\bar{\Sigma}}
e^{i\frac{\tilde{H}_5}{2}}(\bar{z}) \end{eqnarray}
are the supersymmetric charges
and $e^{-\frac{\phi}{2}}$, $e^{i\frac{H_5}{2}}$ are the bosonization
of the superghosts and of the complex fermion associated to the internal 
torus respectively, $S_{\alpha}$ is the space-time spin field operator and 
$\Sigma$ and its complex conjugate are the $K3$ internal spin field
operators; bosonizing the U(1) Cartan current in the SU(2) algebra 
of the internal $N=(4,4)$ SCFT as $J_3=i\sqrt{2}H$, $\Sigma$ can be 
written as $\Sigma=e^{i\frac{\sqrt{2}}{2}H}$. The same thing applies of 
course for the right-moving sector. \\
The 1-loop amplitude involves a sum 
over the torus, Klein bottle, annulus and M\"obius strip surfaces. 
Since the $2g-2$ graviphoton vertex operators are inserted in the (-1) ghost
picture, we need to include in the correlation function $2g-2$ picture 
changing vertex operators. Moreover, due to the +1 charge carried by 
$V_{\gamma}$ in the torus direction, the non-vanishing contribution of 
the picture changing operators will come only from the part $(e^{\phi}e^{-iH_5}
\partial Z_5^+$ + right-moving)\footnote{The discussion done here and in 
what follows also applies to the torus contribution to the amplitude,
reminding that in this case left and right sectors are unrelated, of 
course.}, where $Z_5^+$ is the complex scalar associated to the torus 
direction. The $2g-2$ $\partial Z_5^+$ and $\bar{\partial}Z_5^+$ cannot 
contract with anything in the correlation function, so that only their 
zero modes part give a non-vanishing contribution.
We can use the method of 
images, as described in \cite{bm}, in order to compute the boson and 
fermion propagators on all the surfaces starting from those on the torus.
Bosonizing fermions and superghosts, we can then compute each 
term using the results of \cite{vv}. The important point to note is that 
we have always a cancellation between the contributions of the 
superghosts and the fermions of the torus direction. We are left in this way 
with a sum over the spin structures for the remaining four directions, that
can be related to a triality rotation in the corresponding SO(8) lattice
\cite{gjn}; for
each term, given the structure of the graviton and graviphoton vertex 
operators, the spin-structure dependent part of the amplitude, for generic 
arguments $a_1,a_2$, is always of the form
\be \sum_{g_1,g_2}\sum_{\alpha,\beta} (-)^{4\alpha\beta}\theta^2
\left[^{\textstyle\alpha}_{\textstyle\beta}\right](a_1)
\theta\left[^{\textstyle \alpha+g_1}_{\textstyle \beta+g_2}\right](a_2)
\theta\left[^{\textstyle \alpha-g_1}_{\textstyle \beta-g_2}\right](a_2) \ee
where the first sum is over the twisted sectors of the orbifold (for the 
open string sector $g_1=0$), the second is over the spin structures, the 
first two theta functions refer to the space-time coordinates and the 
remaining two to the internal $K3$ directions. We can now 
perform the spin structure sum using the Riemann identity: 
\begin{eqnarray} &\!&\sum_{g_1,g_2}\sum_{\alpha,\beta} 
(-)^{4\alpha\beta}\theta^2 \left[^{\textstyle \alpha}_{\textstyle \beta}
\right](a_1)\theta\left[^{\textstyle \alpha+g_1}_{\textstyle \beta+g_2}\right]
(a_2)\theta\left[^{\textstyle \alpha-g_1}_{\textstyle \beta-g_2}\right](a_2)
=\nonumber \\
&\!&\sum_{g_1,g_2}\theta\left[^{\textstyle 1/2}_{\textstyle 1/2}\right]
(\tilde{a}_1)\theta\left[^{\textstyle 1/2}_{\textstyle 1/2}\right]
(\tilde{a}_2)\theta\left[^{\textstyle 1/2+g_1}_{\textstyle 1/2+g_2}
\right](0)\theta\left[^{\textstyle 1/2-g_1}_{\textstyle 1/2-g_2}\right](0) 
\end{eqnarray}
where $\tilde{a}_{1,2}=a_1\pm a_2$. 
We can reinterpret this result as an amplitude 
in the odd spin structure
of new vertex operators obtained from the original one
through the SO(8) triality rotation. As shown in \cite{gjn}, the graviton 
vertices are left invariant by the map while the graviphoton 
operators are transformed to:
\be V_{\gamma}(p_1^{\mp})\rightarrow 
\left[(\partial+\bar{\partial})Z_2^{\pm}+ip_1^{\mp}(\psi_1^{\pm}-
\tilde{\psi}_1^{\pm})(\psi_2^{\pm}-\tilde{\psi}_2^{\pm})\right]
e^{ip_1^{\mp}Z_1^{\pm}} \ee
where we did a convenient choice of the kinematical structure. The remarkable 
fact is that the correlation functions now depend only on the space-time 
coordinates, the internal $K3$ part entering through its partition 
function in the odd spin structure. After having performed the 
aforementioned steps and having extracted 
appropriately the kinematical structure, we can define a generating function for the
$F_g$ couplings, whose  expression is given in terms of a simple correlation in a
$\lambda$-perturbed action:
\be\hspace{-.5 cm}
F(\lambda)\equiv\sum_{g=1}^{\infty}g^2 \lambda^{2g}\,F_{g}
=\frac{\lambda^2}{\pi^2}\sum_{\alpha=\rm T,\rm K\atop\rm 
M,\rm A} \int [dM]_{\alpha}\,\sum_{p\in\Gamma}e^{-\frac{\pi [t]|p|^2}{U_2\sqrt{G}}}
C_\alpha^{(0)}([t])
\langle V_g^+ V_g^- e^{-S_0+\tilde{\lambda}S}\rangle_{\alpha} \ee
where $C_{\alpha}^{(0)}$ represent the $K3$ part of the studied indices
and the rest of the 
notation follows \cite{ms}. 
In all the surfaces, because of the four zero modes of the new action,
\be
\langle V_g^+ V_g^- e^{-S_0+\tilde{\lambda}S}\rangle_{\alpha}=\frac{t^2}{2}
\frac{d^2}{d\tilde{\lambda}^2}\langle e^{-S_0+\tilde{\lambda}S}
\rangle_{\alpha} \ee
For $\tilde{\lambda}=0$, i.e. for $F_1$, there is an additional contribution
coming from the the torus, since in this case the action has eight 
zero modes. This further contribution is simply\footnote{Strictly
speaking $F_1$ should be evaluated by a three-point function, as we have done
in subsection 2.1., but the computation reported here reproduces the same 
result.}:
\be F_{1}^{\rm torus}=4\int_{\Gamma}\frac{d^2\tau}{\tau_2}\, C_{\rm T}
\label{tor} \ee
where $\Gamma$ is the fundamental domain of the torus.
We are then left to evaluate determinants of space-time bosons and fermions
and the indices $C_{\alpha}([t])$.
Eq.(\ref{tor}) represents the only non-vanishing 
contribution given by the torus, since the 
corresponding determinant is $\lambda$-independent.
For the other surfaces the determinants can be simply computed using 
the corresponding modes expansion of the fields reported in Appendix A. 
For $n\neq 0$, boson and fermion determinants always
cancel. For $n=0$, in the annulus and M\"obius strip the $\lambda$-dependent 
term in the fermion part of the action drops out, while the bosonic
contribution reduces to:
\be \langle e^{-S_0+\tilde{\lambda}S}\rangle_{\rm A}=
\langle e^{-S_0+\tilde{\lambda}S}\rangle_{\rm M}=
\frac{1}{t^3}\prod_{m=1}^{\infty}\left(1-\frac{\tilde{\lambda}^2}
{m^2}\right)^{-2}=\frac{1}{t^3}\left(\frac{\tilde{\lambda}\pi}
{\sin\tilde{\lambda}\pi}\right)^2 \ee
For the Klein bottle, on the other hand, besides the bosonic contribution there is 
also, for $n=0$, a fermionic contribution for $m=$odd leading to: 
\be \langle e^{-S_0+\tilde{\lambda}S}\rangle_{\rm K}=
\frac{4}{t^3}\left[\prod_{m=1}^{\infty}\left(1-\frac{\tilde{\lambda}^2}
{m^2}\right)^{-2}\right]\left[\prod_{k=0}^{\infty}
\left(1-\frac{4\tilde{\lambda}^2}{(2k+1)^2}\right)^{2}\right]
=\frac{4}{t^3}\left(\frac{\tilde{\lambda}\pi}
{\sin\tilde{\lambda}\pi}\right)^2\!\cos^2\tilde{\lambda}\pi \ee 
where the factor four between the contribution of the Klein bottle 
with that of the annulus and M\"obius strip is due to the 
different modular parameter of the covering tori (see Appendix A). 
Note that the space-time contribution to the amplitude for all the surfaces
is given by the string states with $n=0$, i.e. with oscillation number 
zero. This observation, together with the analysis performed in the last 
section for the internal part, allows us to conclude that only BPS states
are contributing to the considered correlation functions.
Putting all the results together, for large torus 
compactification\footnote{We take this limit simply to perform 
the $\tau_1$ modular integration in the torus contribution.},
we have:
\be F(\lambda)=2\lambda^2\int_0^{\infty}\frac{dt}{t} 
\sum_{p\in\Gamma}e^{-\pi t|p|^2} 
\left[\left(n_H^{total}-n_V^{total}\right)
\frac{d^2}{d\bar{\lambda}^2}\left(\frac{\bar{\lambda}}{\sin\bar{\lambda}}
\right)^2+4n^{closed}_V\right] \ee
with $\bar{\lambda}=\lambda\pi tp/2\sqrt{2U_2}$ and $U_2$ the K\"ahler
class of the $T^2$ torus and where we have used the results of section two:
\bea
\frac{C^{(0)}_{\rm A}+C^{(0)}_{\rm M}}{4}+C^{(0)}_{\rm K}&=&
n^{total}_H-n^{total}_V
\nonumber\\
C^{(0)}_{\rm T}-C^{(0)}_{\rm K}&=&2\,n^{closed}_V\nonumber \eea
Comparing the spectrum in \cite{gj} with the values of $C_{\alpha}^{(0)}$ 
given in Appendix B (table 1), we can explicitly check
that this index reproduces separately for all the sectors the difference 
between the number of
four-dimensional  hyper and vectormultiplets.
These are, of course, particular 
free models with no background fields; in general we will have a mixing 
between the sectors, but such that the total contribution $(C_{\rm A}+C_{\rm 
M})/4+C_{\rm  K}$ will always count the number of BPS (hypers-vectors),
as showed in general in section 2.\\
It is worth while to 
point out that the results obtained here for the $C_{\alpha}^{(0)}$ are 
not in contradiction with what found in \cite{ms} for the 
${\bf Z}_2^A$ model. The values obtained in \cite{ms}, that is 
$C_{\rm T}^{(0)}=8,C_{\rm K}^{(0)}=0,C_{\rm A}^{(0)}
+C_{\rm M}^{(0)}=4\cdot 240$ 
take into account the U(1) anomalies that give 
masses to 16 hypermultiplets in the closed string spectrum and 16 
vectormultiplets to the open one. 

\section{Conclusions}

In this paper we have studied one-loop amplitudes of $N=2$ supersymmetric 
effective actions arising from $K3\times T^2$ compactifications of type I 
string theory.
The important point is the realization of an $N=2$ supersymmetric
index which provides an important information about the BPS spectrum
of the different type I compactifications. Many of the corrections
to chiral terms in these effective actions can be written in terms
of this supersymmetric index, encoding the compactification 
model dependence. Using the superconformal algebras underlying these
compactifications the index is reduced to the counting of BPS states
similar to the one found for the heterotic string \cite{hm}. 
As in that case, we 
proved that this index counts simply the difference between the number
of four-dimensional hyper and vectormultiplets of the corresponding model.\\
Among the one-loop amplitudes related to this index we studied in
detail those corresponding to gravitational higher derivative couplings
$F_g W^{2g}$ for type I compactifications continously connected to  
$K3$ orientifolds. These couplings are completely 
determined by the BPS spectrum of the corresponding model, encoded in the 
aforementioned index.\\ 
These results are independent of any string duality statement. In 
particular all the considered orientifold models besides the $Z_2$
case cannot have a weak heterotic dual in the limit of large torus, 
(where the heterotic winding modes decouple and a weak-weak duality
makes sense) since they contain more than one tensormultiplet
in six dimensions. The structure of the one loop amplitudes we considered
here for type I string compactifications are, however, the same than what
previously found for the heterotic case. This illustrates once more 
how BPS states contributions in correlation functions of different
string theories are organized in an unifyed way
and point out again the importance of these states for a better
understanding of the string dynamics.\\

{\bf{Acknowledgements}}

We thank E. Gava, K.S. Narain, M.H. Sarmadi and
G. Thompson for useful and enlightning discussions.

\vskip 1.5cm 
%\newpage
\section*{Appendix A}
\vskip 0.2 cm
{\em Modes expansions}
\vskip 0.2 cm
\renewcommand{\theequation}{A.\arabic{equation}}
\setcounter{equation}{0}
We present here the modes expansion of the bosonic and fermionic fields 
in the annulus, M\"obius strip and Klein bottle surfaces, needed to 
evaluate the corresponding determinants. We take as fundamental region of 
each surface $0\leq\tau\leq t,0\leq\sigma\leq 1$, where $t$ is the 
corresponding modulus. Following Burgess and Morris \cite{bm}, we can consider
each surface as a torus modded out by a given projection, whose action on the
bosonic and fermionic fields defines the corresponding boundary and crosscap
conditions
\footnote{Note that since 
we need to compute determinants in the odd spin structure, we will 
consider in the following fermionic fields on this spin structure only.}.
The modes expansion for the fields in the anulus is then given by:
\begin{eqnarray}
x_{\rm A}(\tau,\sigma)&=&\sum_{m=-\infty\atop n\geq0}^{+\infty}\alpha_{m,n}
e^{2i\pi m\tau}\cos \pi n\sigma \nonumber \\
\psi_{\rm A}(\tau,\sigma)&=&\sum_{m,n=-\infty}^{+\infty}d_{m,n} 
e^{2i\pi m\tau}e^{i\pi n\sigma} \\
\tilde{\psi}_{\rm A}(\tau,\sigma)&=&\sum_{m,n=-\infty}^{+\infty}d_{m,n} 
e^{2i\pi m\tau}e^{-i\pi n\sigma} \nonumber  \end{eqnarray}
after having extended the fields to the torus 
$0\leq\tau\leq t,0\leq\sigma\leq 2$ and identified 
$x(\tau,\sigma)=x(\tau,2-\sigma),\psi(\tau,\sigma)=\tilde{\psi}
(\tau,2-\sigma)$. For the M\"obius strip we extend the field to the torus
$0\leq\tau\leq 2t,0\leq\sigma\leq 2$ by identifying
$x(\tau,\sigma)=x(1+\tau,1-\sigma),x(\tau,\sigma)=x(\tau,2-\sigma)$,
$\psi(\tau,\sigma)=\tilde{\psi}(1+\tau,1-\sigma),
\psi(\tau,\sigma)=\tilde{\psi}(\tau,2-\sigma)$. Then the mode expansion is:
\begin{eqnarray}
x_{\rm M}(\tau,\sigma)&=&\sum_{m=-\infty\atop n\geq0}^{+\infty}\alpha_{m,n}
e^{i\pi m\tau}\cos \pi n\sigma \ \ \ m+n=even \nonumber \\
\psi_{\rm M}(\tau,\sigma)&=&\sum_{m,n=-\infty}^{+\infty}d_{m,n} 
e^{i\pi m\tau}e^{i\pi n\sigma} \ \ \ m+n=even \\
\tilde{\psi}_{\rm M}(\tau,\sigma)&=&\sum_{m,n=-\infty}^{+\infty}d_{m,n} 
e^{i\pi m\tau}e^{-i\pi n\sigma} \ \ \ m+n=even \nonumber  \end{eqnarray}
The bosonic and fermionic mode expansion in the Klein bottle are:
\begin{eqnarray}
x_{\rm K}(\tau,\sigma)&=&\frac{1}{2}\sum_{m=-\infty\atop 
n\geq0}^{+\infty}\alpha_{m,n}
e^{i\pi m\tau}(e^{2i\pi n\sigma}+(-)^m e^{-2i\pi n\sigma})  \nonumber \\
\psi_{\rm K}(\tau,\sigma)&=&\sum_{m,n=-\infty}^{+\infty}d_{m,n} 
e^{i\pi m\tau}e^{2i\pi n\sigma} \\
\tilde{\psi}_{\rm K}(\tau,\sigma)&=&\sum_{m,n=-\infty}^{+\infty}d_{m,n}
(-)^m e^{i\pi m\tau}e^{-2i\pi n\sigma} \nonumber  \end{eqnarray}
resulting from the identification
$x(\tau,\sigma)=x(1+\tau,1-\sigma),\psi(\tau,\sigma)=
\tilde{\psi}(1+\tau,1-\sigma)$ on the torus
$0\leq\tau\leq 2t,0\leq\sigma\leq 1$. Note that the modular parameter of 
the covering tori for the three surfaces with the aforementioned 
projections is respectively $\tau=it/2,it,2it$.
The factor of two between the annulus and M\"obius strip parameters is due
to the fact that the corresponding tori cover two times the annulus and four
the M\"obius surface.

\section*{Appendix B}
\vskip 0.2 cm
{\em $C_{\alpha}$ for $K3$ orientifolds}
\vskip 0.2 cm
\renewcommand{\theequation}{B.\arabic{equation}}
\renewcommand{\thesection}{B.}
\setcounter{equation}{0}

In this appendix we report the value of the indices $C_{\alpha}^{(0)}$, i.e. of
the indices (\ref{ind1},\ref{ind2}) restricted to the $K3$ part,
for the orientifold models considered in section three.
In order to avoid an heavy notation, we will omit in the following the superscript
$(0)$ in $C_{\alpha}^{(0)}$.\\
$C_{\rm T}={\rm Tr}_{\rm RR}(-)^{F_L+F_R}$ is the Witten index 
\cite{witt}, whose value gives the Euler characteristic of $K3$, that is 
+24, independently of the model\footnote{From now on it will 
be understood that the trace is performed on the $\alpha_N$-invariant 
states on all the sectors, twisted and untwisted.}. The value of 
$C_{\rm K}={\rm Tr}_{\rm RR}\Omega (-)^{F_L+F_R}$\footnote{Note that due 
to the world-sheet parity operator $\Omega$, $F_L=F_R$ so that 
$(-)^{F_L+F_R}$ is completely irrelevant.} can be written for any A-model as
\be C_{\rm K}({\bf Z}_N^A)=-\frac{1}{N}\sum_{k=0}^{N-1}4\sin^2
\frac{2\pi k}{N}+n_{(\frac{N}{2},\frac{N}{2})} \ee where 
the first factor is due to the untwisted sector while
$n_{(\frac{N}{2},\frac{N}{2})}$ is the contribution of the sector twisted
by $\alpha_N^{N/2}$ (when it exists, i.e. for $N\neq 3$) and equals
the number of fixed points invariant under $\alpha_N$ present in that
sector. The other twisted sectors give a vanishing contribution because the
world-sheet parity operator $\Omega$ interchanges sectors twisted by $g$ 
(a generic group element) with the 
ones twisted by $g^{-1}$ and then eigenvalues of $\Omega$ come always in pairs.
For the B-models
$C_{\rm K}={\rm Tr}_{\rm RR}\Omega\alpha_N (-)^{F_L+F_R}$ and its value is
\be C_{\rm K}({\bf Z}_N^B)=-\frac{2}{N}\sum_{k=1}^{N/2}4\sin^2
\frac{2\pi (2k-1)}{N}+n_{(\frac{N}{2},\frac{N}{2})}  \ee
where $n_{(\frac{N}{2},\frac{N}{2})}$ is again the contribution
of the $\alpha_N^{N/2}$-twisted sector, but it now counts the number of fixed points 
invariant under $\alpha_N^2$, weighted by their eigenvalues under $\alpha_N$
\footnote{Remember that in these models there are only the sectors twisted by an 
even number of $\alpha_N$'s, so that $n_{(\frac{N}{2},\frac{N}{2})}=0$ for
the ${\bf Z}_6^B$ model.}.\\ 
\begin{table}[ht]
\begin{center}
\begin{tabular}{||c|c|c||}
\hline {\bf Model} & $(C_{\rm A}+C_{\rm M})/4$ & $C_{\rm K}$ \\ \hline
& 99: -16 & \\ ${\bf Z}_2^A$ & 55: -16 & +16 \\
& 95+59: +256 & \\ \hline
${\bf Z}_3^A$ & 99: -28 & -2 \\ \hline
& 99: -8 & \\ ${\bf Z}_4^A$ & 55: -8 & +8 \\ & 59+95: +128 & \\ \hline
& 99: -20 & \\ ${\bf Z}_6^A$ & 55: -20 & +4 \\ & 59+95: +96 & \\ \hline
${\bf Z}_4^B$ & - & 0 \\ \hline
${\bf Z}_6^B$ & 55: -28 & -2 \\ \hline
\end{tabular}
\caption{\it Values of the indices $C_{\alpha}$ for the various 
surfaces in the open and closed string sectors.} 
\end{center}
\end{table}
Let us now turn our 
attention to the open string indices $C_{\rm A}=
{\rm Tr}_{\rm R}(-)^{F}$
and $C_{\rm M}={\rm Tr}_{\rm R}\,\Omega \,(-)^{F}$, considering separately the
99, 55 and 95+59 sectors. 
Taking into account the results of \cite{gj}
for the open massless spectrum, we can write for all the A-models in the 
99 sector:
\begin{eqnarray} C_{\rm A}^{99}({\bf Z}_N^A)&=&-\frac{1}{N}
\sum_{k=0}^{N-1}4\sin^2\frac{\pi k}{N}({\rm Tr}\gamma_{k,9})^2
\nonumber \\ C_{\rm M}^{99}({\bf Z}_N^A)&=&+\frac{1}{N}
\sum_{k=0}^{N-1}4\sin^2\frac{\pi k}{N}
{\rm Tr}(\gamma_{\Omega_k,9}^{-1}\gamma_{\Omega_k,9}^t)
\end{eqnarray}
following the same notation of \cite{gj}, where the minus sign in 
$C_{\rm A}$ is due to the fermionic charges of the two spin fields in the
Ramond sector. The B-models do not have D9 branes at all. In the 55 sector
\begin{eqnarray} C_{\rm A}^{55}({\bf Z}_N^A)&=&-\frac{1}{N}
\sum_{k=0}^{N-1}4\sin^2\frac{\pi k}{N}({\rm Tr}\gamma_{k,5})^2
\nonumber \\ C_{\rm M}^{55}({\bf Z}_N^A)&=&+\frac{1}{N}
\sum_{k=0}^{N-1}4\cos^2\frac{\pi k}{N}
{\rm Tr}(\gamma_{\Omega_k,5}^{-1}\gamma_{\Omega_k,5}^t)
\end{eqnarray}
where, since $\Omega\psi_0^{3,4}\Omega^{-1}=-\psi_0^{3,4}$ in the 5-sector,
we have $\sin^2\rightarrow\cos^2$ in $C_{\rm M}$. 
For the ${\bf Z}_6^B$ model
\begin{eqnarray} C_{\rm A}^{55}({\bf Z}_6^B)&=&-\frac{2}{6}
\sum_{k=0}^{2}4\sin^2\frac{2\pi k}{6}({\rm Tr}\gamma_{2k,5})^2
\nonumber \\ C_{\rm M}^{55}({\bf Z}_6^B)&=&+\frac{2}{6}
\sum_{k=1}^{3}4\cos^2\frac{\pi (2k-1)}{6}
{\rm Tr}(\gamma_{\Omega_{2k-1},5}^{-1}\gamma_{\Omega_{2k-1},5}^t)
\end{eqnarray}
Finally, in the 95+59 sector:
\be C_{\rm A}^{95+59}({\bf Z}_N^A)=+\frac{2}{N}
\sum_{k=0}^{N-1}({\rm Tr}\gamma_{k,9})({\rm Tr}\gamma_{k,5}) \ee
Given the solution for the matrices $\gamma$'s representing the 
orientifold group \cite{gj}, we can explicitly compute the values of 
these indices for all the models (see table 1). 

\end{document}